\documentclass[11pt,a4paper]{article}

\usepackage[
  a4paper,
  top          = 0.85in,
  left         = 1.5in,
  right        = 1.15in,
  bottom       = 1.5in,
  footskip     = 0.75in,
  marginparwidth = 2.0in
]{geometry}

\usepackage[utf8]{inputenc}
\usepackage[T1]{fontenc}

\usepackage{microtype}
\usepackage{setspace}
\usepackage{ragged2e}
\justifying
\usepackage{csquotes}
\usepackage[left]{lineno}
\usepackage{changepage}

\usepackage{lastpage}

\usepackage[
  backend      = bibtex,
  style        = authoryear-comp,
  sorting      = nyt,
  maxbibnames  = 99,
  maxcitenames = 2,
  uniquelist   = false,
  giveninits   = true,
  natbib       = true,
  dashed       = false,
  isbn         = false,
  url          = false,
  doi          = false,
  eprint       = false
]{biblatex}

\renewbibmacro*{cite:labelyear+extradate}{%
  \iffieldundef{labelyear}
    {}
    {\bibhyperref{\printlabeldateextra}}}

\DeclareFieldFormat[article]{journaltitle}{\textit{#1}}
\DeclareFieldFormat[article]{volume}{#1}
\DeclareFieldFormat[article]{number}{(#1)}
\DeclareFieldFormat[article]{pages}{pp.\addnbspace #1}
\DeclareFieldFormat[article]{eid}{#1}
\DeclareFieldFormat[article]{title}%
  {\textquoteleft\MakeSentenceCase{#1}\textquoteright}
\DeclareFieldFormat{doi}{%
  doi:\space\href{https://doi.org/#1}{\nolinkurl{https://doi.org/#1}}}

\renewbibmacro{in:}{}

\renewbibmacro*{date+extradate}{%
  \iffieldundef{labelyear}
    {}
    {\printtext[parens]{\printlabeldateextra}}}
\renewbibmacro*{author/translator+others}{%
  \ifuseauthor
    {\printnames{author}%
     \iffieldundef{authortype}
       {}
       {\setunit{\addspace}%
        \usebibmacro{authorstrg}}}
    {\usebibmacro{editor/translator+others}}}

\DefineBibliographyStrings{english}{%
  andothers = {\textit{et\addabbrvspace al}\adddot}}

\DeclareBibliographyDriver{article}{%
  \usebibmacro{bibindex}%
  \usebibmacro{begentry}%
  \usebibmacro{author/translator+others}%
  \setunit{\addspace}%
  \usebibmacro{date+extradate}%
  \setunit{\addspace}%
  \printfield{title}%
  \setunit{,\addspace}%
  \printfield{journaltitle}%
  \setunit{,\addspace}%
  \printfield{volume}%
  \printfield{number}%
  \setunit{,\addspace}%
  \iffieldundef{eid}
    {\printfield{pages}}
    {\printfield{eid}}%
  \iffieldundef{doi}
    {}
    {\setunit{.\addspace}\printfield{doi}}%
  \usebibmacro{finentry}}

\usepackage{titlesec}

\renewcommand{\thesection}{\arabic{section}}
\renewcommand{\thesubsection}{\thesection.\arabic{subsection}}
\renewcommand{\thesubsubsection}{\thesubsection.\arabic{subsubsection}}

\titleformat{\section}
  {\normalfont\large\bfseries}{\thesection}{0.5em}{}
\titlespacing*{\section}{0pt}{14pt plus 4pt minus 2pt}{6pt plus 2pt}

\titleformat{\subsection}
  {\normalfont\normalsize\bfseries}{\thesubsection}{0.5em}{}
\titlespacing*{\subsection}{0pt}{10pt plus 3pt minus 1pt}{4pt plus 1pt}

\titleformat{\subsubsection}
  {\normalfont\normalsize\itshape}{\thesubsubsection}{0.5em}{}
\titlespacing*{\subsubsection}{0pt}{8pt plus 2pt minus 1pt}{2pt}

\usepackage{amsmath, amssymb, amsfonts, mathtools}
\usepackage{siunitx}
\usepackage{physics}
\usepackage{upgreek}
\usepackage{gensymb}
\usepackage{bigints}
\usepackage{empheq}

\usepackage{graphicx}
\usepackage{float}
\usepackage{caption}
\usepackage{subcaption}
\usepackage{booktabs}
\usepackage{array}
\usepackage{longtable}
\usepackage{multirow}
\usepackage{tabularx}

\newcolumntype{P}[1]{>{\centering\arraybackslash}m{#1}}
\newcolumntype{R}[1]{>{\raggedright\arraybackslash}p{#1}}
\newcolumntype{L}[1]{>{\raggedleft\arraybackslash}p{#1}}

\usepackage{enumitem}
\setlist[itemize]  {noitemsep, topsep=3pt, parsep=0pt, partopsep=0pt}
\setlist[enumerate]{noitemsep, topsep=3pt, parsep=0pt, partopsep=0pt}

\usepackage{xcolor}
\definecolor{Gray}{gray}{0.25}
\usepackage{tikz}
\usetikzlibrary{arrows.meta, positioning, shapes.geometric}
\usepackage{pgfplots}
\pgfplotsset{compat=1.18}

\PassOptionsToPackage{hyphens}{url}
\usepackage{url}
\usepackage[
  colorlinks = true,
  linkcolor  = blue,
  citecolor  = blue,
  urlcolor   = blue
]{hyperref}
\usepackage{orcidlink}

\usepackage{fancyhdr}
\fancypagestyle{plain}{
  \fancyhf{}
  \lfoot{\scriptsize\sffamily Bolo \& Bagunu~|~{Topological structure of radiation-induced DNA damage}}
  \rfoot{\scriptsize\sffamily\thepage~|~\pageref*{LastPage}}
  
}

\newenvironment{structuredabstract}{%
  \small
  \setstretch{1.1}
}{%
  \par
}

\usepackage[most]{tcolorbox}
\tcbset{
  masterbox/.style={
    enhanced, colback=gray!5, colframe=gray!40,
    boxrule=0.5pt, arc=2pt, left=6pt, right=6pt,
    top=4pt, bottom=4pt
  },
  pipebox/.style={
    enhanced, colback=white, colframe=gray!55,
    boxrule=0.6pt, arc=3pt,
    left=8pt, right=8pt, top=6pt, bottom=6pt,
    fonttitle=\small\bfseries,
    attach boxed title to top left={yshift=-2mm, xshift=6pt},
    boxed title style={colback=black, colframe=gray!55,
      boxrule=0.5pt, arc=2pt, left=4pt, right=4pt}
  }
}

\begin{document}
\thispagestyle{plain}
\vspace*{0.2in}

\begin{flushleft}
{\LARGE\textbf{Topological structure of radiation-induced DNA damage encodes coupled LET--oxygen signatures}}
\\[0.6em]
Renato III Fernan Bolo\textsuperscript{1,*}\,\orcidlink{0009-0000-0878-7096},\quad
Ramon Jose C.\ Bagunu\textsuperscript{1}\,\orcidlink{0009-0002-9306-1971}
\\[0.5em]
{\footnotesize $^1$Department of Physical Sciences and Mathematics,
University of the Philippines Manila, Philippines\\[0.15em]
*Corresponding Author Email: \href{mailto:rfbolo@up.edu.ph}{rfbolo@up.edu.ph}}
\end{flushleft}


\medskip

\begin{structuredabstract}
	\normalsize
  \noindent\textbf{Abstract}

  \medskip

  \noindent
  We present the first nuclear-scale persistent homology and Random Forest classification analysis of radiation-induced DNA double-strand break (DSB) topology across the clinical particle therapy range. Using TOPAS-nBio and the Voxel-Aware Oxygen model, we generated 2,450 simulated nuclei across 49 conditions (seven particle configurations, 0.2--70.7~keV/\textmu{}m; seven oxygen levels, 0.005--21\%~O$_2$) and extracted a 107-feature matrix across seven modalities. DSB topology encodes particle identity, Spread-Out Bragg Peak (SOBP) position, and oxygen tension in a three-tier hierarchy, with fidelity at each tier governed by the physical mechanism controlling it. Particle identity and SOBP position are exactly decodable (balanced accuracy = 1.000). Oxygen-level classification degrades monotonically with LET from 0.517 (electrons) to 0.189 (carbon distal SOBP), with a charge-driven non-monotonicity at the helium-to-carbon transition confirming that atomic number, not LET alone, governs topological discriminability. The joint 49-class task achieves balanced accuracy 0.346, seventeen times above chance. Per-class recall peaks universally at 0.5\%~O$_2$ (0.788--0.976 across all configurations), which is consistent with the OER curve inflection. Topological Summaries (persistent entropy, landscape integrals) dominate oxygen encoding at all LET ($\eta^2_{O_2} =\,$0.300--0.622). A partial-out test reveals two mechanistically separable channels: a count-mediated scale signal ($\eta^2_{O_2}$ survival ratio 0.062) and a count-independent shape signal preserved or enhanced in five of seven configurations (balanced accuracy survival ratio 1.011). Persistent entropy and landscape integrals, as novel radiobiological observables, provide a computational basis for characterizing oxygen-dependent damage topology in hypoxic tumor treatment planning.

\end{structuredabstract}

\vspace{0.5em}

\section{Introduction}
\doublespacing

The oxygen enhancement ratio is the most precisely characterized scalar in radiation biology, and that precision is the measure of what it does not contain. At its most fundamental level, the repair of radiation-induced DNA double-strand breaks (DSBs) is governed by geometry. The cell nucleus maintains a well-defined topological organization through which small charged ions migrate rapidly along thermodynamic potential gradients shaped by chromatin architecture, while large repair complexes move slowly through the same landscape. This geometric organization controls which repair molecules reach which breaks, on what timescale, and by which pathway~\cite{ambrosio2025,caron2020,scully2019}. The OER discards exactly that arrangement.

The spatial distribution of DSBs at the moment of induction is therefore the boundary condition from which repair proceeds, and the topology of breakpoint regions changes in defined patterns that encode repair pathway choice between non-homologous end joining and homologous recombination~\cite{kuntzelmann2026,hahn2021}. How that arrangement varies with irradiation conditions, such as with particle type, linear energy transfer, and oxygen tension, has not been characterized. The damage field has been counted, but its topology has not been asked for.

Linear energy transfer (LET) controls the degree of spatial clustering. At low LET, double-strand breaks are distributed nearly at random across the nuclear volume. At high LET, they concentrate along discrete particle tracks, producing dense, geometrically differentiated clusters whose spatial architecture differs categorically from the low-LET case~\cite{weidner2023,friedrich2018}. Oxygen tension exerts a complementary modulation. The Voxel-Aware Oxygen model (VOxA)~\cite{BoloBagunu2026voxa} demonstrates that at severe hypoxia, DSB retention is suppressed preferentially in low-energy local environments, thus thinning the damage field in a spatially non-uniform manner that depends on the sub-voxel energy landscape left by the incident particle. LET and oxygen therefore control not only the number of breaks but their geometric arrangement, and that arrangement is the object that nuclear repair mechanisms encounter~\cite{schaefer2024}.

Existing parameterizations of the oxygen enhancement ratio (OER)~\cite{grimes2015,scifoni2013} treat oxygen concentration as the sole modulating variable and LET as a fixed scale factor, which collapses both axes onto a single number. While not a calibration limitation, it is an architectural one. At matched LET (e.g., 80~keV/\textmu{}m), helium, carbon, and neon ions produce OER values differing by up to 20\% in the hierarchy $\mathrm{He} < \mathrm{C} < \mathrm{Ne}$~\cite{furusawa2000,strigari2018}. No LET-only model can reproduce this, because track core radius scales with atomic number $Z$ independently of LET, which generates categorically different ionization-density environments whose DSB clustering patterns differ by construction, not by calibration. No volume of measurements can supply the missing particle-identity term, because that term is absent from the model's architecture. DSB topology carries it, while the OER scalar does not.

Persistent homology~\cite{edelsbrunner2002,zomorodian2005} has been applied to sub-resolution single-molecule localization microscopy (SMLM) coordinates of DNA repair proteins within individual foci, where loop and connectivity structure are identified inside repair clusters of $\sim$400--500~nm diameter~\cite{hofmann2018,hausmann2020,hahn2021,kuntzelmann2026,scherthan2023,schaefer2024}. At the nuclear scale, Weidner et al.~\cite{weidner2023} characterized the transition from stochastic to clustered DSB spatial distributions using all-to-all distance histograms and Vietoris-Rips persistent homology, but did not vary oxygen tension. Hu et al.~\cite{hu2025a,hu2025b} examined oxygen as an explicit variable, modelling aerobic and hypoxic conditions through DSB-induction probabilities within a TAD-occupancy framework and establishing that topologically associated DSBs are substantially more lethal than isolated breaks, with incidence ratios across LET and oxygen aligning with the OER trend. However, their framework employs DBSCAN-based cluster counting, which collapses the continuous birth-death diagram to a threshold-dependent count, discarding persistence lifetime as an observable and forgoing the Wasserstein distance structure that enables inter-condition discriminability. The continuous Vietoris-Rips topological analysis of nuclear-scale DSB coordinates across oxygen levels therefore remains a structural gap, not merely an unexplored one.

Nuclear-scale DSB topology encodes particle identity, SOBP position, and oxygen tension in a three-tier hierarchy whose information content and classification fidelity are governed by the physical mechanism that each axis controls. We demonstrate this by applying Vietoris-Rips persistent homology and Random Forest classification to 2,450 simulated nuclei spanning seven particle configurations (0.2--70.7~keV/\textmu{}m) and seven oxygen levels (0.005\%--21\%~O$_2$), generated with TOPAS-nBio~\cite{mcnamara2017,schuemann2019a} and VOxA~\cite{BoloBagunu2026voxa}, constructing a 107-feature matrix across seven modalities.

\section{Materials and Methods}

\subsection{Monte Carlo simulation}

Seven particle configurations ordered by increasing LET span three clinically relevant species at two SOBP positions each, plus a photon surrogate: electrons at 0.2~keV/\textmu{}m representing the secondary electron spectrum of a 6~MV photon beam, protons at the proximal and distal SOBP (4.6 and 8.1~keV/\textmu{}m), helium ions at 10.0 and 30.0~keV/\textmu{}m, and carbon ions at 40.9 and 70.7~keV/\textmu{}m. All normoxic simulations were carried out with TOPAS-nBio~\cite{mcnamara2017,schuemann2019a} using the HalfCylinder chromatin geometry with parameters validated by Bertolet et al.~\cite{bertolet2022}: nucleus radius at 4.65~\textmu{}m, fiber radius at 18.5~nm, fiber length at 120.0~nm, hydration shell thickness at 0.16~nm, and Hilbert-curve spatial arrangement at layer~4 with 30 axial repeats. The \texttt{TsEmDNAPhysics} module governed electromagnetic physics, direct strand-break probability was sampled from a linear model over 5.0--37.5~eV, and histone scavenging was enabled. DSBs were identified by the \texttt{DNADamageNucleusStepByStep} scorer using the 10~base-pair clustering window native to the Standard DNA Damage format~\cite{schuemann2019b}, corresponding to one to two helical turns of the DNA duplex~\cite{cucinotta2024}. DSBs were reconstructed from the SDD output via Hopcroft-Karp maximum bipartite matching, which pairs backbone breaks on opposing strands within the 10~bp window and promotes co-localised DSB pairs to DSB$^{++}$. Each DSB was classified following the Huang et al.~\cite{huang2015} MCDS taxonomy as simple (DSB), DSB$^+$, or DSB$^{++}$; base damages were counted separately. Fifty independent nucleus rotations per configuration provided the biological replication unit at normoxia (21\%~O$_2$), which then yields 350 normoxic nuclei.

\subsection{Hypoxic DSB population generation}

Hypoxic DSB populations were derived from the normoxic simulations using the Voxel-Aware Oxygen model (VOxA)~\cite{BoloBagunu2026voxa}, which models competitive radical chemistry at the sub-millisecond chemical stage. For each DSB, the local energy deposited in a 2$\times$2$\times$2 voxel neighborhood (voxel edge 0.309~\textmu{}m) enters a Michaelis-Menten fixation probability parameterized by the composite kinetic threshold $K_\mathrm{fix} + K_\mathrm{repair} = 0.371\%$~O$_2$ (2.82~mmHg)~\cite{BoloBagunu2026voxa}. DSBs are retained by Bernoulli sampling against this per-DSB probability; the random seed is prefix-derived to guarantee the nested-subset property across oxygen levels, so that every retained DSB at a lower pO$_2$ is also retained at every higher pO$_2$ for the same nucleus. Six hypoxic conditions were generated per configuration at 5.0\%, 2.1\%, 0.5\%, 0.1\%, 0.021\%, and 0.005\%~O$_2$, giving 49 particle-oxygen conditions and 2,450 nuclei in total (50 per condition).

\subsection{Multimodal feature extraction}

Seven feature modalities were computed per nucleus from the retained DSB positions, chromatin coordinates, and voxel energy grids (Table~\ref{tab:modalities}). The complete 107-feature matrix was standardized to zero mean and unit variance per feature before classification.

\begin{table}
\caption{Feature modalities derived per simulated nucleus. $d$: 
number of scalar features per modality.
\texttt{m6} DSB complexity classes (DSB, DSB$^+$, DSB$^{++}$) follow 
the taxonomy of Huang et al.~\cite{huang2015} with the 10~bp clustering 
window justified by Cucinotta~\cite{cucinotta2024}.}
\label{tab:modalities}
\footnotesize
\begin{tabularx}{\textwidth}{@{}lXr@{}}
\toprule
ID / Modality & Representative features & $d$ \\
\midrule
\texttt{m1} Spatial Distribution
  & Nearest-neighbor distances, Ripley $K/L$, radial profile, H0/H1 persistent homology & 33 \\[2pt]
\texttt{m2} Radial Track Structure
  & Off-axis radial fractions, radial entropy, axial-radial ratio, 2D Ripley $K$ & 14 \\[2pt]
\texttt{m3} Local Energy Heterogeneity
  & Voxel energy mean, CV, Gini coefficient, hot-spot fractions, enrichment index & 13 \\[2pt]
\texttt{m4} Dose Distribution
  & Dose percentiles ($D_{10}$, $D_{50}$, $D_{90}$), entropy, CV, high-dose fractions & 10 \\[2pt]
\texttt{m5} Genomic Location
  & Per-chromosome DSB counts, inter-DSB genomic distances, 1D persistence gap statistics & 16 \\[2pt]
\texttt{m6} Damage Complexity
  & Fractions DSB, DSB$^+$, DSB$^{++}$; base damage and backbone strand-break statistics & 11 \\[2pt]
\texttt{m7} Topological Summaries
  & H0/H1 Betti numbers, persistence landscape integrals and peak locations, persistent entropy & 10 \\
\midrule
Total & & 107 \\
\bottomrule
\end{tabularx}
\end{table}

\subsection{Random Forest classification}

Four classification tasks probed distinct axes of information content. Task~1 predicts oxygen level within each particle configuration separately (7-class, one classifier per configuration). Task~2 predicts particle configuration from pooled data (7-class). Task~3 predicts the full 49-class joint condition. Task~4 predicts SOBP position (proximal vs.\ distal) within each particle species. All classifiers used 500 trees~\cite{breiman2001}, balanced class weights, square-root feature subsampling at each split, and the Gini criterion for node splitting. Performance was estimated by stratified 5-fold, 10-repeat cross-validation, and the single-modality ablation used 200 trees and 5-fold, 5-repeat cross-validation. Balanced accuracy (BA), defined as the mean per-class recall, served as the primary metric. Chance level is $1/7 \approx 0.143$ for 7-class tasks and $1/49 \approx 0.020$ for the 49-class task. Feature importance was quantified by permutation importance: the mean decrease in balanced accuracy when each feature is randomly permuted on the held-out set, averaged over five permutation repeats per fold and aggregated across all folds.

\subsection{Topological analysis and partial-out test}

Persistent homology was computed per nucleus from the 3D DSB point
cloud by the Vietoris-Rips filtration~\cite{edelsbrunner2002,bukkuri2021}
implemented in Ripser~\cite{bauer2021}, with a maximum filtration
radius of 9.3~\textmu{}m (the nuclear diameter upper bound), extracting
H0 (connected-component) and H1 (loop) persistence diagrams using
$\mathbb{Z}/2$ coefficients.
Ten \texttt{m7} summary features were derived per diagram pair: the persistence
landscape integral $\int\lambda_1(t)\,\mathrm{d}t$ and peak location $\arg\max_t\,\lambda_1(t)$~\cite{bubenik2015,adams2017}, the persistent entropy $-\sum_i(l_i/L)\log(l_i/L)$~\cite{atienza2020}, whose almost sure convergence for stationary processes was recently established~\cite{thomas2025}, and the mean and variance of the Betti curve $\beta(r)$ computed over a 200-point filtration grid for each of $H_0$ and $H_1$. Pairwise Wasserstein-2 distances between all 2,450 diagrams were computed via Persim~\cite{tralie2018,adams2017}. Within-condition and between-condition distance distributions were summarized by their medians, and the separation ratio (between-condition median divided by within-condition median) quantified topological discriminability across the joint LET-oxygen space.

To decompose the oxygen signal of \texttt{m7} into count-mediated and count-independent components, an ordinary least-squares residualization was applied. The DSB count feature \texttt{m1\_n\_dsbs} was regressed from each of the ten \texttt{m7} features across all 2,450 nuclei via \texttt{numpy.linalg.lstsq}, and Task~1 classifiers were retrained on the resulting residuals using the same 500-tree, 5-fold~$\times$~10-repeat protocol as in Section~2.4.

\section{Results}

Nuclear-scale DSB topology encodes particle identity, SOBP position, and oxygen tension in a three-tier hierarchy whose fidelity at each tier is governed by the physical mechanism that controls it: track geometry for the upper two, radical chemistry for the third.

\subsection{Three-tier classification hierarchy}

DSB topology encodes particle type and SOBP position exactly and oxygen level approximately, with classification difficulty scaling monotonically with LET. Task~2 (7-class particle-type identification) and Task~4 (2-class SOBP position within each species) each achieve BA~=~1.000~$\pm$~0.000 across all cross-validation folds and all species (Table~\ref{tab:task_results}). Task~3 (joint 49-class classification) achieves BA~=~0.346~$\pm$~0.017, 17 times above the 49-class chance level of 0.020.

\begin{table}[h]
\caption{Summary of classification task results. BA: 
balanced accuracy (mean~$\pm$~SD), 5-fold~$\times$~10-repeat 
cross-validation, 500 trees, balanced class weights. Task~4 
reported per species.}
\label{tab:task_results}
\begin{tabularx}{\columnwidth}{@{}cXcl@{}}
\toprule
Task & Target & $N_\mathrm{class}$ & BA (mean $\pm$ SD) \\
\midrule
1 & O$_2$ level, per particle    & 7  & 0.189--0.517 \\
2 & Particle configuration       & 7  & 1.000 $\pm$ 0.000 \\
3 & Joint particle-O$_2$ condition        & 49 & 0.346 $\pm$ 0.017 \\
4 & SOBP position, proton        & 2  & 1.000 $\pm$ 0.000 \\
4 & SOBP position, helium        & 2  & 1.000 $\pm$ 0.000 \\
4 & SOBP position, carbon        & 2  & 1.000 $\pm$ 0.000 \\
\bottomrule
\end{tabularx}
\end{table}

\begin{figure}[!ht][!b]
\includegraphics[width=\textwidth]{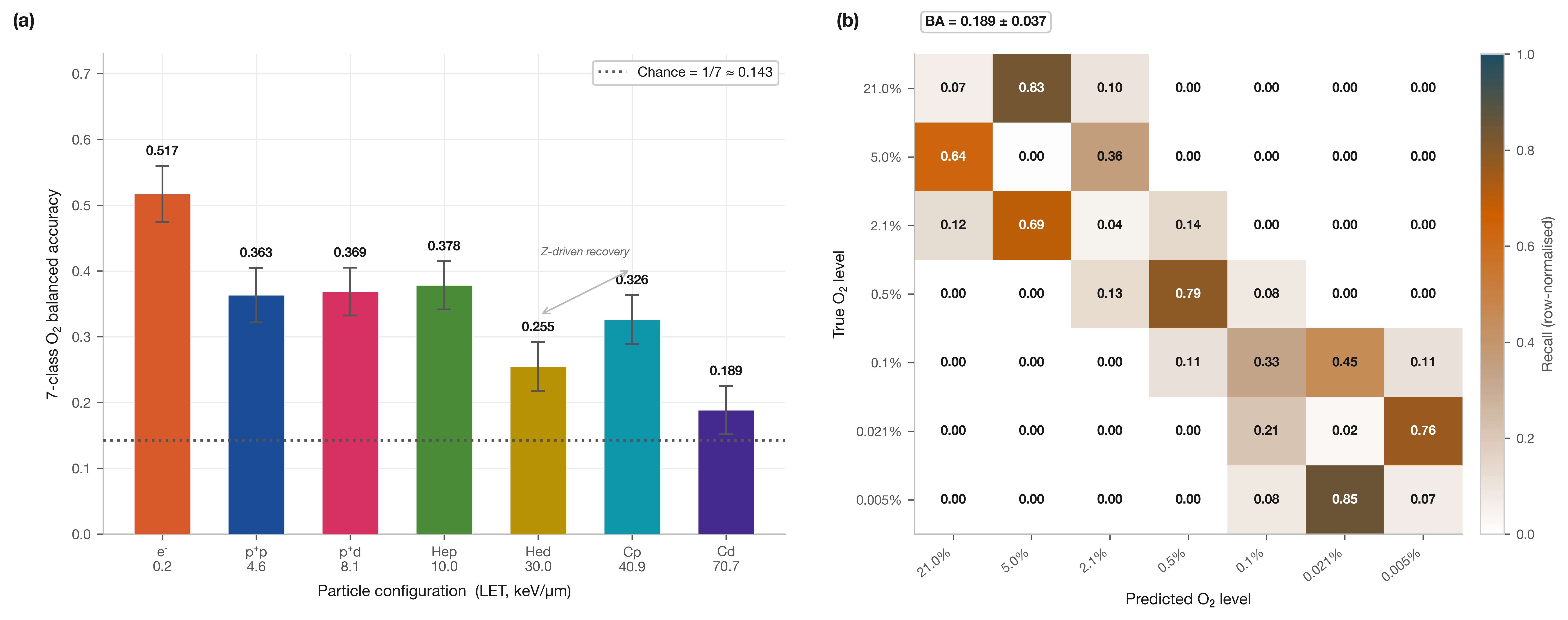}
\caption{Classification of oxygen tension by DSB topology features.
\textbf{a} Task~1 (7-class O$_2$ classification) balanced accuracy (BA)
for each of the seven particle configurations, ordered by LET.
Error bars: $\pm$1~SD over 5-fold~$\times$~10-repeat stratified
cross-validation (500 trees, balanced class weights).
Dotted line: chance level (1/7~$\approx$~0.143).
The annotated bracket marks the helium-distal to carbon-proximal
non-monotonicity ($Z$-driven recovery; see Discussion).
\textbf{b} Row-normalized confusion matrix for the carbon distal SOBP
configuration (LET = 70.7~keV/\textmu{}m, BA~=~0.189~$\pm$~0.037),
the condition with the lowest O$_2$ classification accuracy.}
\label{fig:hierarchy}
\end{figure}

Task~1 oxygen-level classification presents the central problem. Balanced accuracy declines monotonically from BA~=~0.517~$\pm$~0.043 at 0.2~keV/\textmu{}m (electrons) to BA~=~0.189~$\pm$~0.037 at 70.7~keV/\textmu{}m (carbon distal SOBP), with all seven configurations exceeding the chance level of 0.143 (Fig.~\ref{fig:hierarchy}a).

The monotonic descent is interrupted by one systematic inversion: helium distal SOBP (30.0~keV/\textmu{}m, BA~=~0.255) is classified with lower accuracy than carbon proximal SOBP (40.9~keV/\textmu{}m, BA~=~0.326), despite the latter having higher LET. At 30.0~keV/\textmu{}m, helium ($Z$~=~2) produces a diffuse track that distributes DSBs broadly, which generates a topological encoding whose oxygen-dependent variation is less structured than that of the narrower carbon ($Z$~=~6) track at 40.9~keV/\textmu{}m. This concentrates energy in a dense core and produces spatially correlated DSB clusters whose connectivity structure changes distinctly across oxygen levels. The recovery is $Z$-driven because the difference in track geometry accords one particle condition greater topological discriminability at matched LET (see Discussion).

The carbon distal SOBP confusion matrix (Fig.~\ref{fig:hierarchy}b) identifies a universal topological anchor, where the 0.5\%~O$_2$ condition achieves per-class recall within the range 0.788--0.976 across all seven particle configurations. The 0.5\%~O$_2$ level lies immediately above the VOxA composite kinetic threshold $K_\mathrm{fix} + K_\mathrm{repair} = 0.371\%$~O$_2$ (2.82~mmHg)~\cite{BoloBagunu2026voxa}, where the rate of change of DSB retention probability with respect to oxygen tension is largest.

\subsection{Modality hierarchy and functional architecture}

\begin{figure}[!ht]
\includegraphics[width=\textwidth]{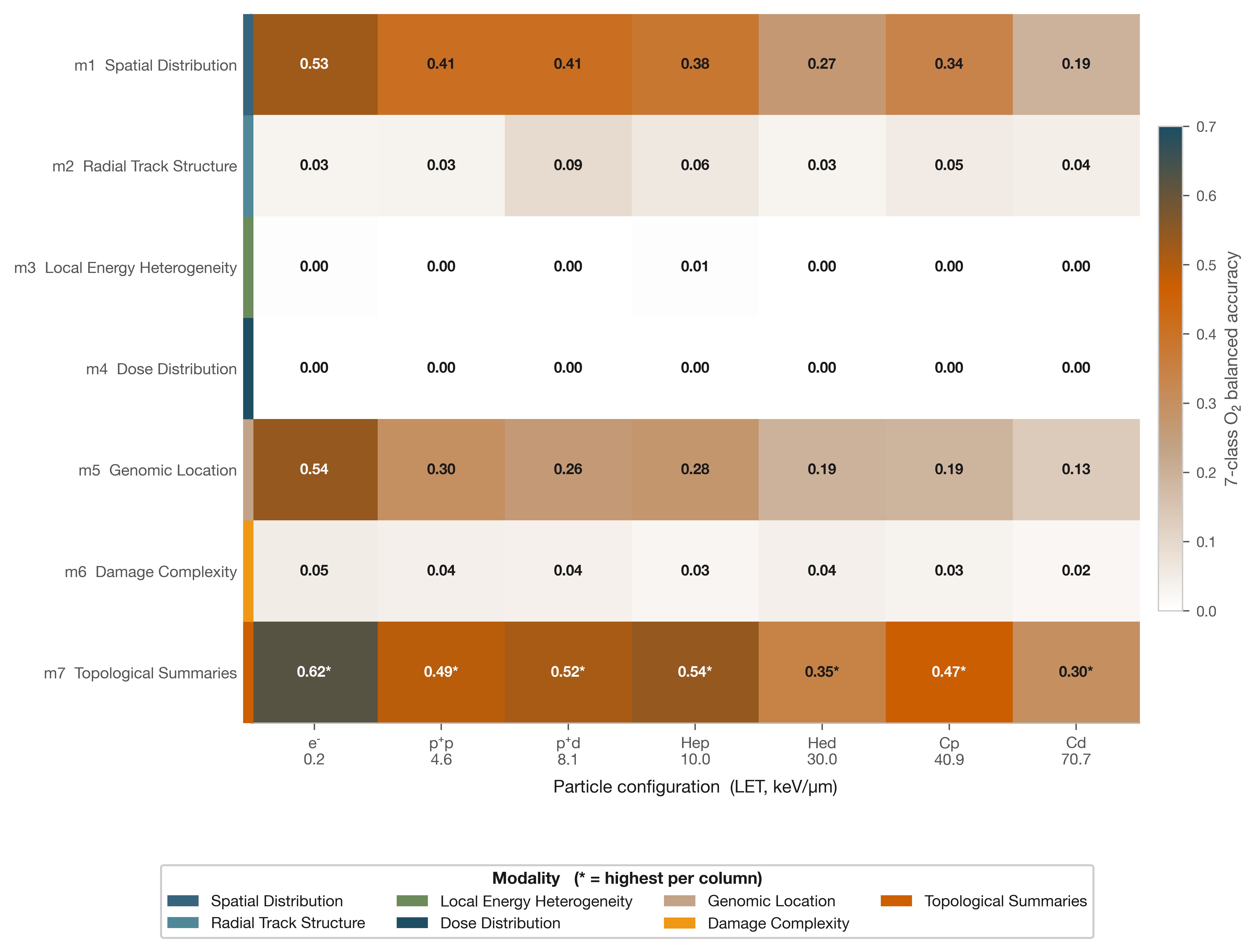}
\caption{Single-modality oxygen-level classification accuracy.
Each cell gives the 7-class balanced accuracy for a Random Forest
trained on one modality alone for one particle configuration
(200 trees, 5-fold~$\times$~5-repeat cross-validation, balanced class
weights).
Asterisk (*): highest-accuracy modality per column.
Color bands on the left margin identify modalities \texttt{m1}--\texttt{m7}. \texttt{m7} Topological Summaries achieves the highest per-column accuracy in
every configuration without exception.}
\label{fig:modality}
\end{figure}

The \texttt{m7} Topological Summaries modality dominates oxygen-level classification across all seven particle configurations. In the single-modality ablation (Fig.~\ref{fig:modality}), \texttt{m7} achieves the highest per-column balanced accuracy for every configuration without exception, ranging from BA~=~0.300 at carbon distal SOBP to BA~=~0.622 at electrons. No other modality achieves this consistency. \texttt{m1} Spatial Distribution and \texttt{m5} Genomic Location are competitive at low LET (\texttt{m1}: BA~=~0.526; \texttt{m5}: BA~=~0.541 for electrons) but degrade more steeply with increasing LET, falling to BA~=~0.194 and BA~=~0.133 respectively at carbon distal SOBP. The \texttt{m3} Local Energy Heterogeneity and \texttt{m4} Dose Distribution modalities produce chance-level oxygen classification across the entire LET range. Voxel energy deposition and macroscopic dose statistics are determined entirely by track structure and carry no oxygen information by construction.

\begin{figure}[!ht]
\includegraphics[width=\textwidth]{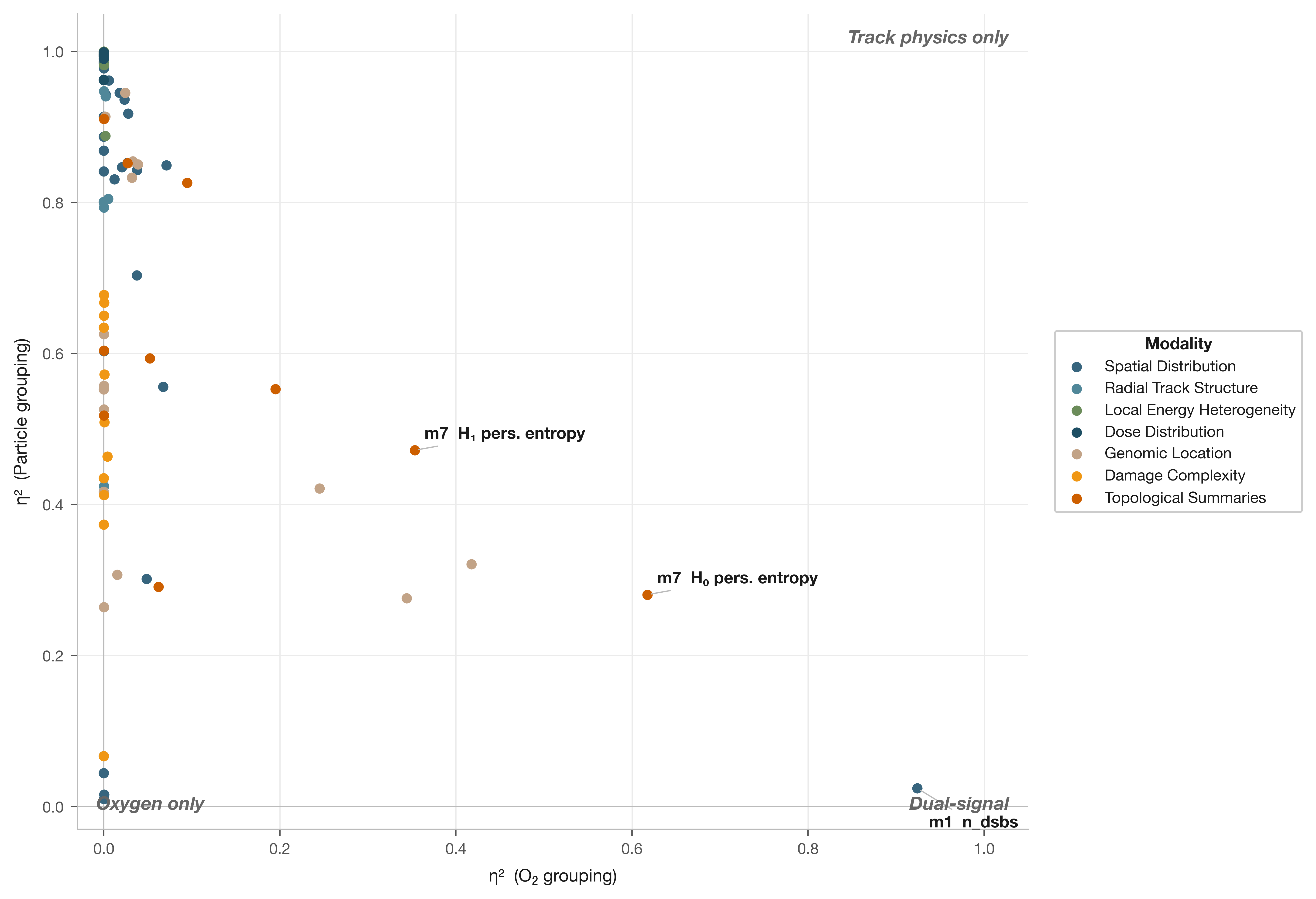}
\caption{Dual-axis scatter of one-way ANOVA effect sizes ($\eta^2$) for
all 107 features.
$x$-axis: $\eta^2$ with O$_2$ level as grouping variable.
$y$-axis: $\eta^2$ with particle type as grouping variable.
Each point represents one feature; color encodes modality.
Three landmark features are labeled.
Upper region: \texttt{m2}, \texttt{m3}, \texttt{m4} features encoding track physics
($\eta^2_\mathrm{particle} \approx 1$, $\eta^2_{O_2} \approx 0$).
Left region: \texttt{m1}, \texttt{m5}, \texttt{m7} oxygen-sensitive features.
Central region: \texttt{m7} persistent entropy features, which respond to both
axes simultaneously.}
\label{fig:scatter}
\end{figure}

The dual-axis effect-size scatter (Fig.~\ref{fig:scatter}) partitions the 107 features into three structural regions. Track-physics features (\texttt{m2}, \texttt{m3}, \texttt{m4}) occupy the high-$\eta^2_\mathrm{particle}$, low-$\eta^2_{O_2}$ region. The most oxygen-sensitive feature, \texttt{m1\_n\_dsbs}, occupies the opposite extreme ($\eta^2_{O_2}$~=~0.924, $\eta^2_\mathrm{particle}$~=~0.024), acting as a pure oxygen counter. The two dominant \texttt{m7} features bridge the central region: H$_0$ persistent entropy ($\eta^2_{O_2}$~=~0.617, $\eta^2_\mathrm{particle}$~=~0.281) and H$_1$ persistent entropy ($\eta^2_{O_2}$~=~0.353, $\eta^2_\mathrm{particle}$~=~0.472) respond to both axes simultaneously. Among \texttt{m5} features, per-chromosome DSB count dispersion ($\eta^2_{O_2}$~=~0.418) and maximum per-chromosome count ($\eta^2_{O_2}$~=~0.344) carry additional oxygen signal independent of 3D spatial topology.

\subsection{Topological discriminability across the LET-oxygen space}

\begin{figure}[!ht]
\includegraphics[width=\textwidth]{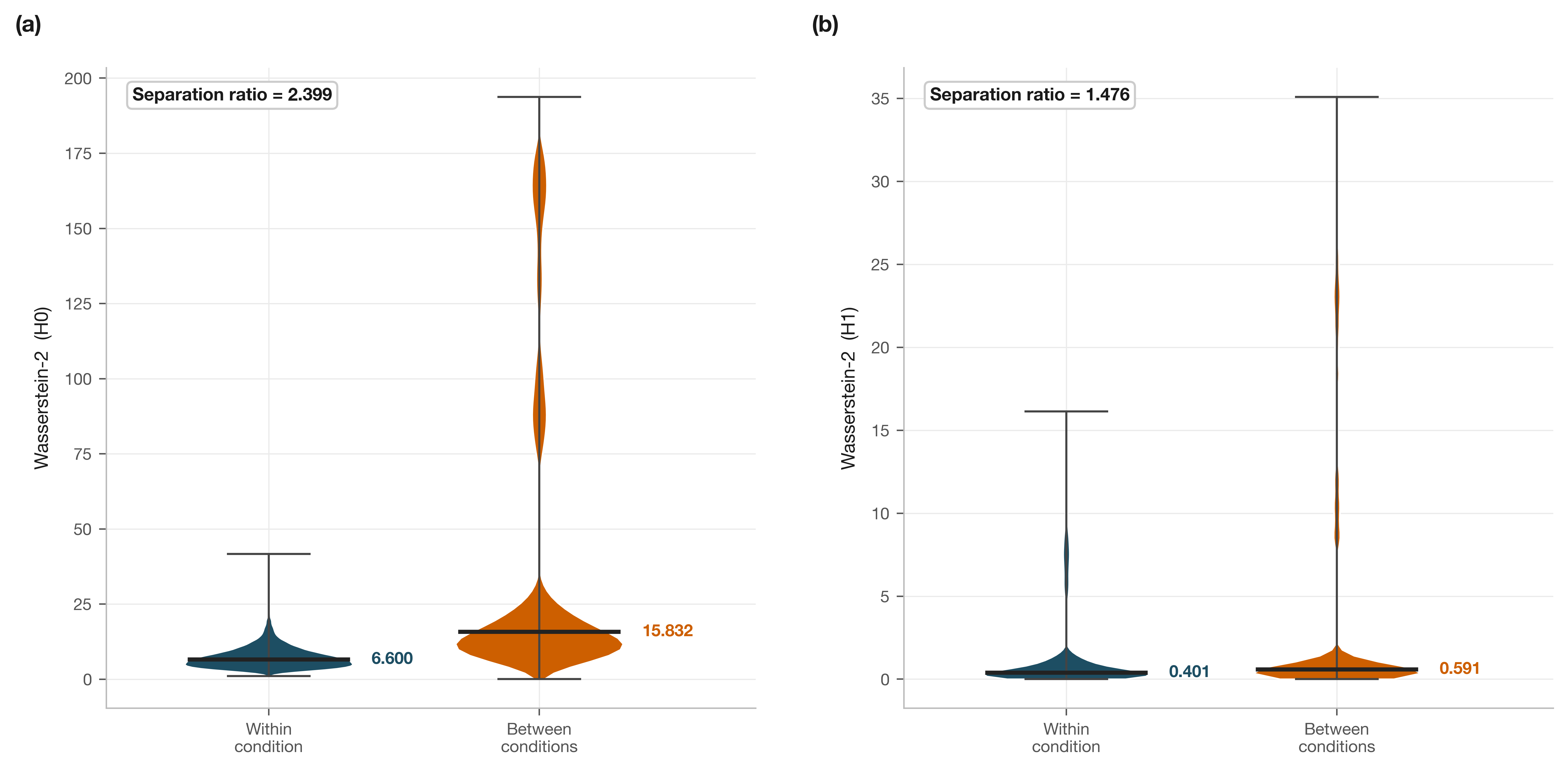}
\caption{Within-condition and between-condition Wasserstein-2 distance
distributions for \textbf{a} H$_0$ and \textbf{b} H$_1$ persistence
diagrams, computed over all 2,450 nucleus pairs
(60,025 within-condition pairs; 2,940,000 between-condition pairs).
Violin plots show the full distribution; bold horizontal bars indicate
medians (values labeled to the right of each violin).
Separation ratio: between-condition median divided by within-condition
median.
The long upper tail in \textbf{a} extends above Wasserstein-2 = 100,
produced by carbon distal SOBP to electron condition pairs, where DSB
count and track geometry differ maximally.}
\label{fig:wasserstein}
\end{figure}

Persistent homology separates particle-oxygen conditions at the level of individual nuclei. H$_0$ Wasserstein-2 distances (Fig.~\ref{fig:wasserstein}a) yield a within-condition median of 6.600 and a between-condition median of 15.832, giving a separation ratio of 2.399. H$_1$ distances (Fig.~\ref{fig:wasserstein}b) yield medians of 0.401 and 0.591 with a separation ratio of 1.476. The H$_0$ ratio exceeds the H$_1$ ratio by a factor of 1.6, consistent with H$_0$ (connected-component lifetime) responding primarily to DSB count variation while H$_1$ (loop persistence) captures subtler changes in spatial arrangement.

\subsection{Dual oxygen-encoding mechanism in \texttt{m7}}

\begin{figure}[!ht]
\includegraphics[width=\textwidth]{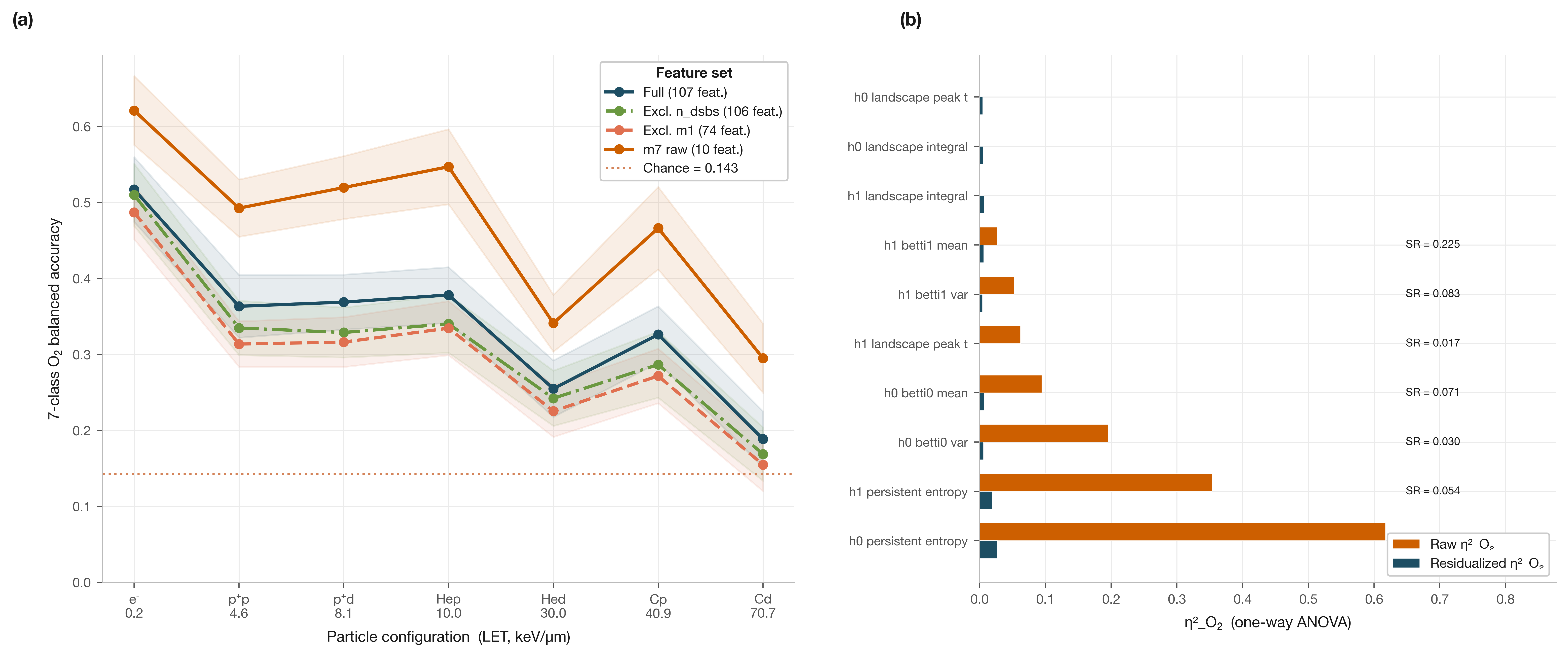}
\caption{Partial-out test characterizing the dual oxygen-encoding
mechanism of \texttt{m7} Topological Summaries.
\textbf{a} 7-class O$_2$ balanced accuracy under five feature-set
conditions: Full (all 107 features), $-n_\mathrm{DSBs}$ (106),
$-$\texttt{m1} (74), \texttt{m7} raw (10), and \texttt{m7} residualized (10 features after OLS
removal of the DSB count feature from each \texttt{m7} feature).
Dotted line: chance level (0.143).
\textbf{b} \texttt{m7} feature-level $\eta^2_{O_2}$ before (raw) and after
(residualized) partialling out DSB count.
SR: survival ratio (residualized / raw).
Together, \textbf{a} and \textbf{b} demonstrate that \texttt{m7} oxygen
information decomposes into a count-mediated scale component
($\eta^2$ collapses after count removal; mean SR~=~0.062) and a
count-independent topological shape component (RF balanced accuracy
preserved or enhanced in five of seven configurations; mean BA
survival ratio 1.011).}
\label{fig:partialout}
\end{figure}

The \texttt{m7} oxygen signal decomposes into two separable components: a count-mediated scale signal and a count-independent shape signal (Fig.~\ref{fig:partialout}).

Count-mediated variance is demonstrated by the $\eta^2_{O_2}$ collapse. H$_0$ persistent entropy falls from $\eta^2_{O_2}$~=~0.617 to 0.027 after OLS removal of \texttt{m1\_n\_dsbs} (survival ratio SR~=~0.044); H$_1$ persistent entropy falls from 0.353 to 0.019 (SR~=~0.054). Averaged over all ten \texttt{m7} features, the $\eta^2_{O_2}$ survival ratio is 0.062: the scale dimension of \texttt{m7} oxygen encoding is almost entirely count-mediated (Fig.~\ref{fig:partialout}b).

The count-independent shape signal is demonstrated by the preservation of classification accuracy after residualization. Classifiers trained on the ten residualized \texttt{m7} features match or exceed their raw counterparts in five of seven particle configurations (proton pSOBP, proton dSOBP, helium dSOBP, carbon dSOBP, and electrons to within noise: $\Delta$BA~=~$-$0.006, below one-tenth of the cross-validation SD). The mean BA survival ratio is 1.011 (Fig.~\ref{fig:partialout}a). The effect is sharpest for proton distal SOBP (BA$_\mathrm{resid}$~=~0.569 vs.\ BA$_\mathrm{raw}$~=~0.520) and carbon distal SOBP (BA$_\mathrm{resid}$~=~0.311 vs.\ BA$_\mathrm{raw}$~=~0.295), the two configurations where high LET most suppresses oxygen-dependent count variation.

\section{Discussion}

\subsection{The three-tier hierarchy as a radiobiological statement}

The classification result states a physical fact about the information content of nuclear-scale DSB topology, not a property of the classifier. Particle identity and SOBP position encoding their full information in the feature matrix (BA~=~1.000) reflects that at matched dose, a carbon ion ($Z$~=~6) and a helium ion ($Z$~=~2) deposit energy in categorically different geometric configurations, imprinting categorically different DSB clustering patterns on the nucleus. The monotonic LET-dependence of oxygen classification follows from the same physics by a complementary mechanism. High-LET particles shift DSB induction toward the direct ionization pathway~\cite{hirayama2009,strigari2018}. Because oxygen fixation competes specifically with thiol-mediated restitution of indirect radical damage~\cite{grimes2015,chapman1979}, a larger direct-damage fraction means a smaller oxygen-sensitive fraction and reduced topological contrast between normoxic and hypoxic DSB populations. No feature modality, including topology, can recover the oxygen signal that LET has physically eliminated.

\subsection{The Z-driven inversion}

The helium-distal to carbon-proximal non-monotonicity is not an artifact of sample size or classifier variance. At 30.0~keV/\textmu{}m, helium ($Z$~=~2) produces a diffuse track that distributes DSBs broadly across the nucleus, generating an inter-DSB distance distribution whose oxygen-dependent variation is topologically unstructured. At 40.9~keV/\textmu{}m, carbon ($Z$~=~6) concentrates energy in a nanometer-scale core, producing a sharp spatial clustering of DSBs whose connectivity structure changes distinctly across oxygen levels, evidenced by the $\sim$50~nm peak in inter-DSB distance histograms~\cite{friedrich2018,weidner2023}. The recovery is $Z$-driven: structured carbon-ion geometry is more topologically discriminable than diffuse helium-ion geometry at comparable LET. This inversion will present an independent test case for downstream analyses operating through different geometric frameworks.

\subsection{The 0.5\% oxygen anchor}

The universal discriminability of the 0.5\%~O$_2$ condition (per-class recall 0.788--0.976 across all particle configurations) is anchored in the VOxA fixation kinetics. The VOxA composite threshold $K_\mathrm{fix} + K_\mathrm{repair} = 0.371\%$~O$_2$ (2.82~mmHg)~\cite{BoloBagunu2026voxa} recovers the classical K-value of competitive radical kinetics, locating the OER curve inflection immediately below the 0.5\%~O$_2$ experimental level, where the partial derivative of DSB retention probability with respect to oxygen tension is largest and topological contrast between adjacent oxygen levels is greatest~\cite{alper1956,grimes2015}. Clinically, this range corresponds to moderate hypoxia, above the HIF-1$\alpha$ stabilization floor but below well-perfused tumor normoxia, where radiosensitization by supplemental oxygenation has the most steeply dose-modifying potential~\cite{sokol2023,tinganelli2015}. The computational result therefore converges with classical radiobiological first principles.

\subsection{Topological summaries as oxygen sensors}

\texttt{m7}'s dominance across all seven particle configurations (Fig.~\ref{fig:modality}) establishes persistent homology as the most informative single-modality oxygen sensor in this framework. H$_0$ persistent entropy captures this mechanism. Hypoxia preferentially retains DSBs in the highest-energy local environments, producing a smaller, spatially less diverse population. The connectivity structure is more uniform, in the sense of lower persistent entropy, than the normoxic population.

The partial-out result (Fig.~\ref{fig:partialout}) decomposes this into two separable encoding channels. The dominant scale channel, in which persistent entropy amplitude tracks DSB count ($\eta^2_{O_2}$ survival ratio 0.044), is the primary but count-confounded signal. The residual shape channel, in which the relative geometric arrangement of DSBs encodes oxygen independently of their number (BA survival ratio 1.011, enhanced in five of seven configurations), is mechanistically more fundamental. It persists once the scale information is removed.

The shape channel is the component most directly relevant to future experimental validation. When persistent entropy and landscape integrals are computed from SMLM repair-protein localizations, fluorophore labelling efficiency introduces uncalibrated count uncertainty. This confounds the scale signal while leaving the shape signal, the rank-order relationship across conditions, intact. One topological constraint governs the choice of validation criterion. The Vietoris-Rips (VR) filtration used here and the \v{C}ech-based filtrations employed in intra-focus SMLM pipelines obey the interleaving relationship $\mathcal{C}_\alpha \subseteq \mathrm{VR}_\alpha \subseteq \mathcal{C}_{\sqrt{2}\alpha}$ in Euclidean space~\cite{edelsbrunner2002}. Therefore, absolute filtration values are not directly comparable across the two frameworks, as Vietoris-Rips complexes fill topological cycles at systematically lower radii than their \v{C}ech counterparts. The appropriate validation criterion is therefore Spearman rank correlation between simulation-derived and SMLM-derived topological features across matched LET-oxygen conditions, not absolute value agreement, as rank-order is preserved under the monotonic scale shifts that the VR/\v{C}ech interleaving introduces.

\subsection{Scale distinction from intra-focus persistent homology}

The Hausmann group's intra-focus PH program~\cite{hofmann2018,hausmann2020,hahn2021,kuntzelmann2026,scherthan2023,schaefer2024} applies Vietoris-Rips filtration to SMLM localizations within individual repair clusters. This characterizes sub-cluster connectivity and loop structure of the repair apparatus at diameters of $\sim$400--500~nm and filtration radii $\leq$160~nm. The present analysis operates at the nuclear scale, where filtration to the full nuclear DSB field with inter-break distances spans hundreds of nanometers to several micrometers using simulation-derived break coordinates rather than fluorescence localizations. These are not the same measurement at different magnifications. The intra-focus program asks how the repair machinery organizes within a single focus, while the nuclear-scale program asks how the precursor damage field is distributed across the nucleus as a function of irradiation conditions. The two questions are orthogonal: one concerns the response topology, the other the stimulus topology.

The distinction carries a practical consequence for validation design. Intra-focus topology characterizes the repair apparatus at later timepoints post-irradiation, after foci have assembled. Nuclear-scale topology here characterizes the precursor DSB field at chemical equilibrium ($<$1~ms). A future unified analysis would ask specifically whether nuclear-scale topological features at $t$~=~0 predict the intra-focus topological structure that emerges at early post-irradiation timepoints, providing a causal link between damage topology and repair-complex organization. The observables derived here, particularly persistent entropy and landscape integrals computed from simulation coordinates, are simultaneously computable from experimental SMLM localizations of repair proteins, where Spearman rank correlation across matched LET-oxygen conditions provides the appropriate validation criterion, since absolute filtration values are not directly comparable across the two frameworks given the VR/\v{C}ech interleaving in Euclidean space~\cite{edelsbrunner2002}. Neither framework has previously examined oxygen as a variable. This paper and the intra-focus program together occupy non-overlapping regions of the spatial and temporal parameter space, a circumstance that makes their eventual synthesis, not competition, the productive scientific direction.

\subsection{Limitations}

Four limitations bound the scope of the present findings. First, all simulations use a single generic nucleus geometry (HalfCylinder, radius 4.65~\textmu{}m); cell-type-specific chromatin organization is not modeled. Second, the analysis characterizes the precursor DSB population at chemical equilibrium (sub-millisecond); repair dynamics, chromatin mobility, and DSB clustering at later timepoints are outside this scope. Third, the stability of persistent entropy $H_p = -\sum_i(l_i/L)\log(l_i/L)$ as a functional of the persistence diagram has not been established analytically. The standard diagram stability theorem~\cite{edelsbrunner2002} guarantees Lipschitz continuity of diagrams under the bottleneck distance with respect to $L^\infty$ perturbations of the generating function, but the log-normalisation in $H_p$ is non-linear and does not inherit this guarantee directly. The entropy is used here as a discriminative observable, validated empirically by its classification performance and Wasserstein separability, rather than as a provably stable invariant. Fourth, the bridge-observable prediction, that nuclear-scale topological features computed from simulated DSB coordinates will rank-order experimental SMLM cluster topology at early post-irradiation timepoints, remains to be tested and is the target of future experimental work.

\section{Conclusions}

Radiation-induced DSB topology encodes irradiation conditions in a hierarchical structure that directly reflects the physical mechanisms controlling each variable. Particle identity and SOBP position impose categorical differences in track geometry that are exactly readable from the DSB field. Oxygen tension modulates the radical-chemistry fraction of that field in a way that degrades monotonically with LET but never vanishes entirely, as confirmed by the $Z$-driven non-monotonicity at the helium-to-carbon transition, which establishes atomic number, not LET alone, as the governing variable for topological discriminability across particle species.

The most forward-relevant finding is the dual decomposition of the \texttt{m7} oxygen signal. The count-independent shape channel, preserved or enhanced in five of seven particle configurations after DSB count removal, is mechanistically more fundamental than the dominant scale channel. It persists once scale information is removed, making it robust to the count uncertainty that fluorophore labelling efficiency introduces in experimental imaging. H$_0$ persistent entropy ($\eta^2_{O_2}$~=~0.617) and landscape integrals are, to our knowledge, the first topological observables computed from radiation damage data at the nuclear scale and across a clinically relevant LET-oxygen grid.

These results carry direct implications for radiological physics and technology. The 0.5\%~O$_2$ topological anchor, where per-class recall is highest across all particle configurations, corresponds to the moderate-hypoxia range most relevant to LET painting strategies in carbon-ion and helium-ion therapy~\cite{bassler2014,sokol2023}. The observables introduced here, which are persistent entropy and landscape integrals, are simultaneously computable from mechanistic simulations and from experimental single-molecule localization coordinates of DNA repair proteins. This makes the nuclear-scale topological characterization established here a natural reference framework for geometrical analyses of repair focus architecture at later post-irradiation timepoints, where the precursor damage topology identified in this work can be tested as a predictor of the repair organization that follows. Embedding such a framework in treatment-planning geometry requires establishing the quantitative link between simulated damage topology and experimental repair topology across matched LET-oxygen conditions, which defines the scope of future experimental endeavors and computational bridging.

\section*{Funding}
\addcontentsline{toc}{section}{Acknowledgements}

This work received no external funding. All computations were performed on the corresponding author's local workstation.

\textit{Author contributions (CRediT taxonomy).}
R.I.F.B.: Conceptualization, Methodology, Software, Validation, Formal analysis, Investigation, Data curation, Visualization, Writing – original draft, Writing  – review \& editing;
R.J.C.B.: Conceptualization, Methodology, Investigation, Supervision, Project administration, Writing – review \& editing.

The authors declare no conflicts of interest.

\section*{Data Availability Statement}
\addcontentsline{toc}{section}{Data Availability Statement}

The simulation pipeline, feature extraction scripts, and analysis code are available at \url{https://doi.org/10.5281/zenodo.20174484}.

\section*{Other declarations}
\addcontentsline{toc}{section}{Data Availability Statement}

A preliminary version of this work will be presented as an oral contribution at the joint Southeast Asia Conference on Medical Physics (SEACOMP) 2026 and 7th Philippine Conference on Medical Physics (PCMP), June 2026.

\nolinenumbers
\printbibliography[
  heading = bibintoc,
  title   = {References}
]

@article{cucinotta2024,
  author  = {Cucinotta, Francis A.},
  title   = {Modeling clustered {DNA} damage by ionizing radiation using multinomial damage probabilities and energy imparted spectra},
  journal = {International Journal of Molecular Sciences},
  volume  = {25},
  number  = {23},
  pages   = {12532},
  year    = {2024},
  doi     = {10.3390/ijms252312532}
}

@article{sokol2023,
  author  = {Sokol, Olga and Durante, Marco},
  title   = {Carbon ions for hypoxic tumors: Are we making the most of them?},
  journal = {Cancers},
  volume  = {15},
  number  = {18},
  pages   = {4494},
  year    = {2023},
  doi     = {10.3390/cancers15184494}
}

@article{tinganelli2015,
  author  = {Tinganelli, W. and Durante, M. and Hirayama, R. and Kr{\"a}mer, M. and Maier, A. and Kraft-Weyrather, W. and Furusawa, Y. and Friedrich, T. and Scifoni, E.},
  title   = {Kill-painting of hypoxic tumours in charged particle therapy},
  journal = {Scientific Reports},
  volume  = {5},
  pages   = {17016},
  year    = {2015},
  doi     = {10.1038/srep17016}
}

@article{bassler2014,
  author  = {Bassler, Niels and Toftegaard, Jakob and L\"uhr, Armin and S\o{}rensen, Brita Singers and Scifoni, Emanuele and Kr\"amer, Michael and J\"akel, Oliver and Mortensen, Lise Saks\o{} and Overgaard, Jens and Petersen, J\o{}rgen B.},
  title   = {{LET}-painting increases tumour control probability in hypoxic tumours},
  journal = {Acta Oncologica},
  volume  = {53},
  number  = {1},
  pages   = {25--32},
  year    = {2014},
  doi     = {10.3109/0284186X.2013.832835}
}

@article{alper1956,
  author  = {Alper, Tikvah and Howard-Flanders, Paul},
  title   = {Role of oxygen in modifying the radiosensitivity of {E.~coli} {B}},
  journal = {Nature},
  volume  = {178},
  number  = {4540},
  pages   = {978--979},
  year    = {1956},
  doi     = {10.1038/178978a0}
}

@article{furusawa2000,
  author  = {Furusawa, Yoshiya and Fukutsu, Koichi and Aoki, Masami and Itsukaichi, Hiroshi and Eguchi-Kasai, Kiyomi and Ohara, Hitoshi and Yatagai, Fumio and Kanai, Tatsuaki and Ando, Koichi},
  title   = {Inactivation of aerobic and hypoxic cells from three different cell lines by accelerated heavy ions},
  journal = {Radiation Research},
  volume  = {154},
  number  = {5},
  pages   = {485--496},
  year    = {2000},
  doi     = {10.1667/0033-7587(2000)154[0485:IOAAHC]2.0.CO;2}
}

@article{grimes2015,
  author  = {Grimes, D. Robert and Partridge, M.},
  title   = {A mechanistic investigation of the oxygen fixation hypothesis and oxygen enhancement ratio},
  journal = {Biomedical Physics {\&} Engineering Express},
  volume  = {1},
  number  = {4},
  pages   = {045209},
  year    = {2015},
  doi     = {10.1088/2057-1976/1/4/045209}
}

@article{scifoni2013,
  author  = {Scifoni, E. and Tinganelli, W. and Weyrather, W. K.
             and Durante, M. and Maier, A. and Kr{\"a}mer, M.},
  title   = {Including oxygen enhancement ratio in ion beam treatment planning: {Model} implementation and experimental verification},
  journal = {Physics in Medicine {\&} Biology},
  volume  = {58},
  number  = {11},
  pages   = {3871--3895},
  year    = {2013},
  doi     = {10.1088/0031-9155/58/11/3871}
}

@article{strigari2018,
  author  = {Strigari, L. and Torriani, F. and Manganaro, L. and Inaniwa, T. and Dalmasso, F. and Cirio, R. and Attili, A.},
  title   = {Tumour control in ion beam radiotherapy with different ions in presence of hypoxia: {A}n oxygen enhancement ratio model based on the microdosimetric kinetic model},
  journal = {Physics in Medicine and Biology},
  year    = {2018},
  volume  = {63},
  number  = {6},
  eid  	  = {065012},
  doi     = {10.1088/1361-6560/aa89ae}
}

@article{chapman1979,
  author  = {Chapman, J. D.},
  title   = {Hypoxic sensitizers---implications for radiation therapy},
  journal = {New England Journal of Medicine},
  volume  = {301},
  number  = {26},
  pages   = {1429--1432},
  year    = {1979},
  doi     = {10.1056/NEJM197912273012606}
}

@article{huang2015,
  author = {Huang, Y. W. and Pan, C. Y. and Hsiao, Y. Y. and Chao, T. C. and Lee, C. C. and Tung, C. J.},
  title = {Monte Carlo simulations of the relative biological effectiveness for {DNA} double strand breaks from 300 {MeV} u$^{-1}$ carbon-ion beams},
  journal = {Physics in Medicine and Biology},
  volume  = {60},
  number  = {15},
  pages   = {5995--6012},
  year    = {2015},
  doi     = {10.1088/0031-9155/60/15/5995}
}

@article{edelsbrunner2002,
  author  = {Herbert Edelsbrunner and David Letscher and Afra Zomorodian},
  title   = {Topological persistence and simplification},
  journal = {Discrete {\&} Computational Geometry},
  volume  = {28},
  number  = {4},
  pages   = {511--533},
  year    = {2002},
  doi     = {10.1007/s00454-002-2885-2}
}

@article{zomorodian2005,
  author  = {Zomorodian, Afra and Carlsson, Gunnar},
  title   = {Computing persistent homology},
  journal = {Discrete \& Computational Geometry},
  volume  = {33},
  pages   = {249--274},
  year    = {2005},
  doi     = {10.1007/s00454-004-1146-y}
}

@article{bauer2021,
  author  = {Bauer, Ulrich},
  title   = {Ripser: {Efficient} computation of {Vietoris--Rips} persistence barcodes},
  journal = {Journal of Applied and Computational Topology},
  volume  = {5},
  pages   = {391--423},
  year    = {2021},
  doi     = {10.1007/s41468-021-00071-5}
}

@article{tralie2018,
  author  = {Tralie, Christopher and Saul, Nathaniel and Bar-On, Rann},
  title   = {{Ripser.py}: A lean persistent homology library for {Python}},
  journal = {Journal of Open Source Software},
  volume  = {3},
  number  = {29},
  pages   = {925},
  year    = {2018},
  doi     = {10.21105/joss.00925}
}

@article{adams2017,
  author  = {Adams, Henry and Chepushtanova, Sofya and Emerson, Tegan and Hanson, Eric and Kirby, Michael and Motta, Francis and Neville, Rachel and Peterson, Chris and Shipman, Patrick and Ziegelmeier, Lori},
  title   = {Persistence Images: A Stable Vector Representation of Persistent Homology},
  journal = {Journal of Machine Learning Research},
  volume  = {18},
  pages   = {1--35},
  year    = {2017},
  url     = {https://dl.acm.org/doi/abs/10.5555/3122009.3122017}
}

@article{thomas2025,
  author  = {Thomas, A. M.},
  title   = {Convergence of persistence diagrams for discrete time stationary processes},
  journal = {Journal of Applied and Computational Topology},
  volume  = {9},
  eid     = {14},
  year    = {2025},
  doi     = {10.1007/s41468-025-00211-1}
}

@article{bukkuri2021,
  author  = {Bukkuri, Abhinav and Andor, Noemi and Darcy, Isabelle K.},
  title   = {Applications of topological data analysis in oncology},
  journal = {Frontiers in Artificial Intelligence},
  volume  = {4},
  pages   = {659037},
  year    = {2021},
  doi     = {10.3389/frai.2021.659037}
}

@article{hofmann2018,
  author  = {Hofmann, Alexander and Krufczik, Matthias
             and Heermann, Dieter W. and Hausmann, Michael},
  title   = {Using persistent homology as a new approach for super-resolution localization microscopy data analysis and classification of {$\gamma$H2AX} foci/clusters},
  journal = {International Journal of Molecular Sciences},
  volume  = {19},
  number  = {8},
  pages   = {2263},
  year    = {2018},
  doi     = {10.3390/ijms19082263}
}

@article{hausmann2020,
  author  = {Hausmann, Michael and Neitzel, Charlotte and Bobkova, Elizaveta and Nagel, David and Hofmann, Andreas and Chramko, Tatyana and Smirnova, Elena and Kope{\v{c}}n{\'a}, Olga and Pag{\'a}{\v{c}}ov{\'a}, Eva and Boreyko, Alla and others},
  title   = {Single molecule localization microscopy analyses of {DNA}-repair foci and clusters detected along particle damage tracks},
  journal = {Frontiers in Physics},
  volume  = {8},
  pages   = {578662},
  year    = {2020},
  doi     = {10.3389/fphy.2020.578662}
}

@article{hahn2021,
  author  = {Hannes Hahn and Charlotte Neitzel and Olga Kope{\v{c}}n{\'a} and Dieter W. Heermann and Martin Falk and Michael Hausmann},
  title   = {Topological analysis of {$\gamma$H2AX} and {MRE11} clusters detected by localization microscopy during {X}-ray-induced {DNA} double-strand break repair},
  journal = {Cancers},
  volume  = {13},
  number  = {21},
  pages   = {5561},
  year    = {2021},
  doi     = {10.3390/cancers13215561}
}

@article{kuntzelmann2026,
  author  = {K{\"u}ntzelmann, Kim Annabel and Pardo, Laura Rozo and Sch{\"a}fer, Myriam and Weidner, Jonas and Falkova, Iva and Toufar, Jiri and Toufarova, Lucie and Bestvater, Felix and Hausmann, Michael and Falk, Martin},
  title   = {Nanoscale topology of {$\gamma$H2AX} and {53BP1} foci in {U87} cancer cells and normal {NHDF} after high-{LET} radiation-induced {DSB} repair},
  journal = {Nanoscale},
  volume  = {18},
  number  = {11},
  pages   = {4399--4414},
  year    = {2026},
  doi     = {10.1039/d5nr05100b}
}

@article{scherthan2023,
  author  = {Scherthan, Heidemarie and Geiger, Beate and Ridinger, Daniel and M{\"u}ller, Julia and Riccobono, Daniela and Bestvater, Felix and Port, Monika and Hausmann, Michael},
  title   = {Nano-architecture of persistent focal {DNA} damage regions in the minipig epidermis weeks after acute {$\gamma$}-irradiation},
  journal = {Biomolecules},
  volume  = {13},
  number  = {10},
  pages   = {1518},
  year    = {2023},
  doi     = {10.3390/biom13101518}
}

@article{weidner2023,
  author  = {Weidner, Jonas and Neitzel, Charlotte and Gote, Martin and Deck, Jeanette and K{\"u}ntzelmann, Kim and Pilarczyk, G{\"o}tz and Falk, Martin and Hausmann, Michael},
  title   = {Advanced image-free analysis of the nano-organization of chromatin and other biomolecules by single molecule localization microscopy ({SMLM})},
  journal = {Computational and Structural Biotechnology Journal},
  volume  = {21},
  pages   = {2018--2034},
  year    = {2023},
  doi     = {10.1016/j.csbj.2023.03.009}
}

@article{friedrich2018,
  author  = {T. Friedrich and K. Ilicic and C. Greubel and S. Girst and J. Reindl and M. Sammer and B. Schwarz and C. Siebenwirth and D. W. M. Walsh and T. E. Schmid and M. Scholz and G. Dollinger},
  title   = {{DNA} damage interactions on both nanometer and micrometer scale determine overall cellular damage},
  journal = {Scientific Reports},
  volume  = {8},
  number  = {1},
  pages   = {16063},
  year    = {2018},
  doi     = {10.1038/s41598-018-34323-9}
}

@article{hu2025a,
  author  = {Hu, Ankang and Zhou, Wanyi and Luo, Xiyu and Qiu, Rui and Li, Junli},
  title   = {Correlation between {DNA} double-strand break distribution in {3D} genome and ionizing radiation-induced cell death},
  journal = {Radiation Research},
  year    = {2025},
  volume  = {203},
  number  = {6},
  pages   = {421--432},
  doi     = {10.1667/RADE-24-00277.1}
}

@article{hu2025b,
  author   = {Hu, Ankang and Zhou, Wanyi and Luo, Xiyu and Qiu, Rui and Li, Junli},
  title    = {Impact of oxygen on {DNA} damage distribution in {3D} genome and its correlation to oxygen enhancement ratio after high-{LET} irradiation},
  journal  = {Radiation Research},
  year     = {2025},
  doi      = {10.1667/RADE-25-00093.1}
}

@article{mcnamara2017,
  author  = {McNamara, Aidan and Geng, Chen and Turner, Ross and Mendez, Jose R. and Perl, Joseph and Held, Kathryn and Faddegon, Bruce and Paganetti, Harald and Schuemann, Jan},
  title   = {Validation of the radiobiology toolkit {TOPAS-nBio} in simple {DNA} geometries},
  journal = {Physica Medica},
  volume  = {33},
  pages   = {207--215},
  year    = {2017},
  doi     = {10.1016/j.ejmp.2016.12.010}
}

@article{schuemann2019a,
  author  = {Schuemann, J. and McNamara, A. L. and Ramos-Méndez, J. and Perl, J. and Held, K. D. and Paganetti, H. and others},
  title   = {{TOPAS-nBio}: {A}n Extension to the {TOPAS} simulation toolkit for cellular and sub-cellular radiobiology},
  journal = {Radiation Research},
  year    = {2019},
  volume  = {191},
  number  = {2},
  pages   = {125--138},
  doi     = {10.1667/RR15226.1}
}

@article{schuemann2019b,
  author  = {Schuemann, J. and McNamara, A. L. and Warmenhoven, J. W. and Henthorn, N. T. and Kirkby, K. J. and Merchant, M. J. and Ingram, S. and Paganetti, H. and Held, K. D. and Ramos-Mendez, J. and Faddegon, B. and Perl, J. and Goodhead, D. T. and Plante, I. and Rabus, H. and Nettelbeck, H. and Friedland, W. and Kundr{\'a}t, P. and Ottolenghi, A. and Baiocco, G. and McMahon, S. J.},
  title   = {A new standard {DNA} damage ({SDD}) data format},
  journal = {Radiation Research},
  year    = {2019},
  volume  = {191},
  number  = {1},
  pages   = {76--92},
  doi     = {10.1667/RR15209.1}
}

@article{bertolet2022,
  author  = {Bertolet, Alexander and Ramos-M{\'e}ndez, Jose and McNamara, Aidan and Yoo, Donguk and Ingram, Samuel and Henthorn, Nicholas and Warmenhoven, John W. and Faddegon, Bruce and Merchant, Michael and McMahon, Stephen J. and Paganetti, Harald and Schuemann, Jan},
  title   = {Impact of {DNA} geometry and scoring on {Monte Carlo} track-structure simulations of initial radiation-induced damage},
  journal = {Radiation Research},
  volume  = {198},
  number  = {3},
  pages   = {207--220},
  year    = {2022},
  doi     = {10.1667/RADE-21-00179.1}
}

@article{hirayama2009,
  author  = {Hirayama, Ryoichi and Ito, Atsushi and Tomita, Masanori and Tsukada, Teruyo and Yatagai, Fumio and Noguchi, Miho and Matsumoto, Yoshitaka and Kase, Yuki and Ando, Koichi and Okayasu, Ryuichi and Furusawa, Yoshiya},
  title   = {Contributions of direct and indirect actions in cell killing by high-{LET} radiations},
  journal = {Radiation Research},
  volume  = {171},
  number  = {2},
  pages   = {212--218},
  year    = {2009},
  doi     = {10.1667/RR1490.1}
}

@article{breiman2001,
  author  = {Breiman, Leo},
  title   = {Random forests},
  journal = {Machine Learning},
  volume  = {45},
  number  = {1},
  pages   = {5--32},
  year    = {2001},
  doi     = {10.1023/A:1010933404324}
}

@article{BoloBagunu2026voxa,
  author  = {Bolo, Renato III Fernan and Bagunu, Ramon Jose C.},
  title   = {Voxel-aware oxygen kinetics resolves radiation-induced {DNA} damage retention across {LET}-oxygen conditions in particle therapy},
  journal = {arXiv},
  year    = {2026},
  doi	  = {10.48550/arXiv.2605.12558}
}

@article{ambrosio2025,
  author  = {Ambrosio, Susanna and Noviello, Anna and Di Fusco, Giovanni and Gorini, Francesca and Piscone, Anna and Amente, Stefano and Majello, Barbara},
  title   = {Interplay and dynamics of chromatin architecture and {DNA} damage response: An overview},
  journal = {Cancers},
  volume  = {17},
  number  = {6},
  pages   = {949},
  year    = {2025},
  doi     = {10.3390/cancers17060949}
}

@article{caron2020,
  author  = {Caron, Pierre and Polo, Sophie E.},
  title   = {Reshaping chromatin architecture around {DNA} breaks},
  journal = {Trends in Biochemical Sciences},
  volume  = {45},
  number  = {3},
  pages   = {177--178},
  year    = {2020},
  doi     = {10.1016/j.tibs.2019.12.001}
}

@article{scully2019,
  author  = {Scully, Ralph and Panday, Arvind and Elango, Rajula and Willis, Nicholas A.},
  title   = {{DNA} double strand break repair pathway choice in somatic mammalian cells},
  journal = {Nature Reviews Molecular Cell Biology},
  volume  = {20},
  number  = {11},
  pages   = {698--714},
  year    = {2019},
  doi     = {10.1038/s41580-019-0152-0}
}

@article{schaefer2024,
  author  = {Schäfer, Myriam and Hildenbrand, Georg and Hausmann, Michael},
  title   = {Impact of gold nanoparticles and ionizing radiation on whole chromatin organization as detected by single-molecule localization microscopy},
  journal = {International Journal of Molecular Sciences},
  volume  = {25},
  number  = {23},
  pages   = {12843},
  year    = {2024},
  doi     = {10.3390/ijms252312843}
}

@article{atienza2020,
	author	= {Nieves Atienza and Rocio Gonzalez-Díaz and Manuel Soriano-Trigueros},
	title	= {On the stability of persistent entropy and new summary functions for topological data analysis},
	journal = {Pattern Recognition},
	volume	= {107},
	pages	= {107509},
	year	= {2020},
	doi		= {10.1016/j.patcog.2020.107509}
}

@article{bubenik2015,
	author	= {Bubenik, Peter},
	title	= {Statistical topological data analysis using persistence landscapes},
	journal	= {The Journal of Machine Learning Research},
	volume	= {16},
	number	= {1},
	pages	= {77--102},
	year	= {2015},
	url		= {https://dl.acm.org/doi/10.5555/2789272.2789275}
}

\pagebreak

\appendix

\section{Complete partial-out balanced accuracy table}
\label{app:partialout}

Table~\ref{tab:partialout_full} gives the Task~1 balanced accuracy (mean~$\pm$~SD) for all seven particle configurations under each of the five feature-set conditions examined in Section~3.4. Condition labels: \textit{Full}~=~all 107 features; $-n_\mathrm{DSBs}$~=~106 features (\texttt{m1\_n\_dsbs} removed); $-$\texttt{m1}~=~74 features (entire \texttt{m1} modality removed); \texttt{m7}\,raw~=~10 \texttt{m7} features unmodified; \texttt{m7}\,resid.~=~10 \texttt{m7} features after OLS removal of \texttt{m1\_n\_dsbs}. The survival ratio SR is computed as BA(\texttt{m7}\,resid.) / BA(\texttt{m7}\,raw) per configuration. Values consistent with Fig.~\ref{fig:partialout}a.

\begin{table}[!ht]
\caption{Task~1 seven-class O$_2$ balanced accuracy (mean~$\pm$~SD)
for all seven particle configurations under five feature-set conditions.
500 trees, balanced class weights, 5-fold~$\times$~10-repeat stratified
cross-validation.
SR: BA survival ratio (\texttt{m7}\,resid. / \texttt{m7}\,raw).
Dagger ($\dagger$): electron configuration treated as within noise
($\Delta\mathrm{BA} = -0.006$, below one-tenth of the cross-validation SD).}
\label{tab:partialout_full}
\scriptsize
\begin{tabularx}{\textwidth}{@{}Xrrrrrc@{}}
\toprule
Particle (LET, keV/\textmu{}m)
  & Full (107)
  & $-n_\mathrm{DSBs}$ (106)
  & $-$\texttt{m1} (74)
  & \texttt{m7} raw (10)
  & \texttt{m7} resid. (10)
  & SR \\
\midrule
e$^-$ mono (0.2)
  & $0.517 \pm 0.043$
  & $0.510 \pm 0.041$
  & $0.487 \pm 0.035$
  & $0.621 \pm 0.045$
  & $0.615 \pm 0.047$$^{\dagger}$
  & 0.990 \\[2pt]
p$^+$ pSOBP (4.6)
  & $0.363 \pm 0.041$
  & $0.335 \pm 0.036$
  & $0.314 \pm 0.030$
  & $0.493 \pm 0.038$
  & $0.504 \pm 0.039$
  & 1.022 \\[2pt]
p$^+$ dSOBP (8.1)
  & $0.369 \pm 0.036$
  & $0.329 \pm 0.033$
  & $0.316 \pm 0.033$
  & $0.520 \pm 0.042$
  & $0.569 \pm 0.044$
  & 1.094 \\[2pt]
He pSOBP (10.0)
  & $0.378 \pm 0.037$
  & $0.340 \pm 0.038$
  & $0.335 \pm 0.036$
  & $0.547 \pm 0.049$
  & $0.517 \pm 0.047$
  & 0.945 \\[2pt]
He dSOBP (30.0)
  & $0.255 \pm 0.037$
  & $0.242 \pm 0.036$
  & $0.225 \pm 0.034$
  & $0.341 \pm 0.037$
  & $0.364 \pm 0.038$
  & 1.066 \\[2pt]
C pSOBP (40.9)
  & $0.326 \pm 0.037$
  & $0.287 \pm 0.044$
  & $0.272 \pm 0.036$
  & $0.466 \pm 0.054$
  & $0.441 \pm 0.042$
  & 0.946 \\[2pt]
C dSOBP (70.7)
  & $0.189 \pm 0.037$
  & $0.169 \pm 0.035$
  & $0.155 \pm 0.035$
  & $0.295 \pm 0.046$
  & $0.311 \pm 0.041$
  & 1.053 \\
\midrule
Mean
  & $0.342$
  & $0.316$
  & $0.301$
  & $0.469$
  & $0.474$
  & 1.011 \\
\bottomrule
\end{tabularx}
\end{table}

\vfill
\pagebreak

\section{TOPAS-nBio simulation parameters}
\label{app:topas}

Table~\ref{tab:topas_params} lists the complete \texttt{DNADamageNucleusStepByStep} scorer parameters used in all simulations, taken verbatim from the TOPAS-nBio parameter files. Geometry parameters follow Bertolet et al.~\cite{bertolet2022} and are given in Section~2.1. The beam source was a pencil beam (Gaussian spread $\sigma_{xy}$~=~0.05~\textmu{}m, cutoff 0.5~\textmu{}m; angular spread $\sigma_\theta$~=~0.005$^\circ$) directed along the $-z$ axis at a source-to-nucleus distance of 5~\textmu{}m. Particle-specific primary energies were selected to reproduce the dose-averaged LET values listed in Section~2.1.

\begin{table}[!ht]
\caption{TOPAS-nBio \texttt{DNADamageNucleusStepByStep} scorer
parameters applied uniformly across all seven particle configurations.
Parameter values are taken verbatim from the simulation parameter
files; see Bertolet et al.~\cite{bertolet2022} for the biological
basis of the HalfCylinder-specific scavenging and indirect-damage
probabilities.}
\label{tab:topas_params}
\scriptsize
\begin{tabularx}{\textwidth}{@{}llX@{}}
\toprule
Parameter & Value & Description \\
\midrule
\multicolumn{3}{@{}l}{\textit{Physics and geometry}} \\[2pt]
Physics module & TsEmDNAPhysics & EM physics list \\
Nucleus radius & 4.65~\textmu{}m & HalfCylinder scoring volume \\
\multicolumn{3}{@{}l}{\textit{Direct damage model}} \\[2pt]
UseLinearProbabilityForDirectDamage & true & Linear probability in eV window \\
LowerLimitForLinearProbabilityFunction & 5.0~eV & Lower threshold \\
UpperLimitForLinearProbabilityFunction & 37.5~eV & Upper threshold \\
ScoreDirectDamages & true & \\
ScoreQuasiDirectDamages & true & \\
\multicolumn{3}{@{}l}{\textit{Indirect damage model}} \\[2pt]
ScoreIndirectDamages & true & \\
AlwaysScavengeSpeciesInDNAComponents & false & \\
ProbabilityOfScavengingInBackbone & 0.0585 & Bertolet et al. Table~1 \\
ProbabilityOfScavengingInBase & 1.0 & \\
ProbabilityOfIndirectDamageToBackbone & 0.55 & \\
ProbabilityOfIndirectDamageToBase & 1.0 & \\
ProbabilityOfChargeTransferFromHydrationShellToBackbone & 0.3333 & \\
ScavengeInHistones & true & Histone scavenging enabled \\
\multicolumn{3}{@{}l}{\textit{DSB clustering and output}} \\[2pt]
MaximumBasePairDistanceToConsiderDSB & 10~bp & Clustering window \\
BreakDownOutputPerDamageOrigin & true & Direct/indirect breakdown \\
MinimalSDDOutput & false & Full SDD fields written \\
IncludeDSBDamageSitesOnlyinSDD & false & All damage sites in SDD \\
WriteCSVOutputWithAllDamageSpecification & true & Per-break CSV output \\
\multicolumn{3}{@{}l}{\textit{Parallel world voxel scorer}} \\[2pt]
EnergyDeposit and DoseToMedium scorers & 60$^3$ grid & $18.56~\mu\mathrm{m}^3$ box,
  $\Delta x = 0.309$~\textmu{}m \\
\bottomrule
\end{tabularx}
\end{table}

\end{document}